\documentstyle[12pt,epsf]{article}
\def\beq{\begin{equation}}
\def\eeq{\end{equation}}
\def\beqa{\begin{eqnarray}}
\def\eeqa{\end{eqnarray}}
\def\dgs{\delta_{GS}}
\def\dgst{\tilde{\delta}_{GS}}
\def\p{\partial}
\def\t{\tilde}
\def\zii{Z_2 \times Z_6}
\def\ziii{Z_3 \times Z_6}
\def\bmat{\begin{array}}
\def\emat{\end{array}}

\newcommand\mfrac[2]{\mbox{$\frac{#1}{#2}$}}

\begin{document}
\title{The Supersymmetric Particle Spectrum in Orbifold Compactifications of 
String Theory}
\author{A. Love and P. Stadler\thanks{P.Stadler@rhbnc.ac.uk} \\ 
\\Department of Physics\\
Royal Holloway\\
University of London\\Egham, Surrey, TW20 0EX, UK.}
\date{July 1997}
\maketitle

\begin{abstract}
\sloppy
The supersymmetry breaking parameters and the resulting supersymmetric particle
spectrum are studied in orbifold compactifications of string theory under the
assumption that unification of gauge coupling constants at about $10^{16}$ GeV
is a consequence of large moduli dependent string loop threshold corrections.
The effect on the spectrum of various assumptions as to the modular weights of 
the states, the values of the Green-Schwarz parameter, $\dgs$, the origin of 
the $\mu$ parameter and the moduli dependence of Yukawa couplings is discussed.
The effect of radiative corrections to the effective potential is also 
considered.
\end{abstract}
\fussy

\section{Introduction}
In a generic supergravity theory, the soft supersymmetry breaking scalar 
masses, gaugino masses and $A$ and $B$ terms are free parameters. On the other
hand if the supergravity theory is the low energy limit of an orbifold 
compactificaion of the heterotic string then these parameters are calculable in
principle \cite{cvetic,ibanez,carlos,brig,brig2}
since string theory has only one free parameter, namely the 
string scale. However, the soft supersymmetry breaking parameters depend on the
moduli of the orbifold model (including the dilaton expectation value) and the
values of the moduli cannot be determined without a detailed knowledge of the
non-peturbative superpotential probably responsible for supersymmetry breaking.
One possible approach \cite{brig} is to accept for the time being that we lack
this 
detailed knowledge and to absorb this uncertainty into an angle, $\theta$, 
which is a measure of the relative size of the auxiliary fields for the 
dilaton and the overall $T$ modulus (with the vacuum energy taken to be zero).

In such an approach \cite{brig} the supersymmetric particle spectrum has been 
derived
as a function of this angle $\theta$ when unification of gauge coupling 
constants at $2 \times 10^{16}$ GeV is due to large moduli dependent string
loop corrections and also when it is due to extra matter states close to the
unification scale. Here, we shall explore the robustness of the qualitative 
features of the spectrum obtained in the former case when various assumptions
about the orbifold model are varied. In particular, we shall consider the 
effect of making one or more of the following changes to the assumptions in 
Brignole {\em et al}.\ \cite{brig}:
\newline
{\bf a.} Using the modular weights allowed \cite{cvetic,bl1} for states in the
twisted sectors of 
those abelian orbifolds which possess three $N=2$ moduli, $T_i$. Then it is 
possible to adopt a single overall modulus model with $T_1=T_2=T_3=T$ as in 
ref.\cite{brig} with all three moduli on the same footing if, as is the case 
in 
gaugino condensate models, only the $N=2$ moduli occur in the non-perturbative
superpotential. The possible choices of modular weights are then further 
restricted by requiring that the string loop threshold corrections to the 
gauge coupling constants allow unification of all three observable sector 
gauge coupling constants at a single energy scale. The value of the overall
modulus $T$ is determined by requiring that this energy scale is $2 \times 
10^{16}$ GeV.  This is an alternative to choosing the simplest set of modular
weights \cite{brig} which will achieve the gauge coupling unification without 
reference
to any particular orbifold.
\newline
{\bf b.} Adopting the values for the Green-Schwarz parameters, $\dgs$, 
suggested by the above orbifold models.
\newline
{\bf c.} Taking account of the possible moduli dependence of the Yukawa 
couplings when all three states are in twisted sectors of the orbifold.
\newline
{\bf d.} In the case that the $\mu$ parameter originates from K\"ahler 
potential mixing, using the moduli dependence of $\mu$ suggested by the 
discussion of ref.\cite{antoniadis} rather than taking $\mu$ to be moduli 
independent, and in addition,
\newline
{\bf e.} taking account of the radiative corrections to the tree level 
effective potential in calculating the Higgs scalar expectaton values $v_1$
and $v_2$ which affect the supersymmetric particle spectrum and also in 
calculating the Higgs scalar masses \cite{ellis,ellis2}.

We do not consider the effect of more than one independent modulus expectation
value which has been considered elsewhere \cite{brig2}, nor do we consider the 
M-theory regime of strong ten dimensional string coupling \cite{horava} for 
which gauge 
coupling constant unification at the `observed' energy scale may occur without
large string loop threshold corrections if there is a large eleventh dimension
\cite{witten}.

The organisation of the paper is as follows. In section 2 all possible 
choices of modular weights for the standard model states in abelian orbifold
compactifications with three $N=2$ moduli $T_i$ are obtained. The choice is 
restricted by demanding consistency with gauge coupling constant unification
with $T_1=T_2=T_3=T$. The corresponding value of $T$ is also given.
In section 3 the soft supersymmetry breaking terms are presented as functions
of the overall modulus $T$ and the angle $\theta$ introduced in ref.\cite{brig}
. In 
section 4 the relevant renormalisation group equations for the running of the
coupling constants and soft supersymmetry breaking parameters from the 
unification scale to the electroweak scale are displayed and the strategy for 
choosing the various string theoretic parameters and ensuring the correct 
electroweak breaking scale is discussed. In section 5 the resulting 
supersymmetric particle spectrum is explored, including the effect of 
radiative corrections to the effective potential. Finally, in section 6 we 
present our conclusions and make comparisons with the work of ref.\cite{brig}.

\section{Choices of modular weights}
As will be seen in section 3, the values of the soft supersymmetry breaking
parameters at the string scale depend on the modular weights of the matter 
states \cite{cvetic,ibanez,carlos,brig,brig2}. Let us first establish our 
conventions. In general we shall write 
the K\"ahler potential $K$ to quadratic order in the matter fields in the 
form
\beq
K=-\ln{Y}-\sum_i\ln{(T_i + \bar{T}_i)}+\sum_{\alpha}\tilde{K}_{\alpha} 
|\phi_\alpha|^2 +(Z\phi_1 \phi_2 + h.c.) \label{K}
\eeq
with
\beqa
Y&=&S+\bar{S}-\sum_i \delta_i \ln{(T_i + \bar{T}_i)} \\
\delta_i &=& \frac{\dgs^i}{8 \pi^2}
\eeqa
and
\beq
\tilde{K}_\alpha=\prod_i (T_i + \bar{T}_i)^{n^i_\alpha} \label{Kprod}
\eeq

In (\ref{K})-(\ref{Kprod}) any $U$ moduli associated with $Z_2$ planes are 
included as additional $T_i$ moduli, $\dgs^i$ are Green-Schwarz parameters, 
$\phi_{\alpha}$ are matter fields and the $Z \phi_1 \phi_2$ term is present 
when the orbifold has a $Z_2$ plane (ie. when the action of the point group 
in that plane is as $Z_2$). The matter fields $\phi_1$ and $\phi_2$ are 
untwisted states associated with the $T$ and $U$ moduli for the $Z_2$ plane.
The powers $n^i_{\alpha}$ are the modular weights for the matter fields 
$\phi_\alpha$.

In the case of a single overall modulus 
\beq
T=T_1=T_2=T_3 
\eeq
these expressions reduce to
\beq
K=-\ln{Y}-3\ln{(T + \bar{T})}+\sum_{\alpha}\tilde{K}_{\alpha} 
|\phi_\alpha|^2 +(Z\phi_1 \phi_2 + h.c.)
\eeq
with
\beq
Y=S+\bar{S}-\dgst\ln{(T+\bar{T})}
\eeq
where
\beqa
\dgs&=&\sum_i \dgs^i \label{gs}\\
\dgst&=& \frac{\dgs}{8 \pi^2} \label{gst}
\eeqa
and
\beq
\tilde{K}_\alpha=(T+\bar{T})^{n_\alpha}
\eeq
with overall modular weights 
\beq
n_\alpha=\sum_i n^i_\alpha \label{n}\ \ .
\eeq

The only abelian orbifolds that possess three $N=2$ moduli $T_i$ are $\zii$ 
and $\ziii$, the former orbifold having in addition a 
single $U$ modulus. All possible modular weights for massless matter states
in the twisted (and untwisted) sectors of abelian orbifolds can be determined 
using the approach of refs. \cite{ibanez} and \cite{bl1}. For $\zii$ the 
allowed modular weights are 
\beq
\mbox{(Q, u, e) : } n_\alpha=0,-1,-2 \label{mw1}
\eeq
and
\beq
\mbox{(L, d, H) : } n_\alpha=+1,0,-1,-2,-3  \label{mw2}
\eeq
where Q, L, and H denote quark, lepton and Higgs $SU(2)_L$ doublets and 
u, d and e denote quark and lepton singlets. For $\ziii$, the 
possible modular weights are
\beq
\mbox{(Q, u, e) : } n_\alpha=0,-1,-2 \label{mw3}
\eeq
and
\beq
\mbox{(L, d, H) : } n_\alpha=+1,0,-1,-2,-3,-4 \label{mw4}
\eeq
For a single overall modulus $T$ the conditions for unification of the 
$SU(3)_C \times SU(2)_L \times U(1)$ gauge coupling constants $g_3$, $g_2$ and
$\tilde{g}_1$ at a scale less than $10^{18}$ GeV may be taken to be 
\cite{deren,ibanez2,ibanez}
\beq
\frac{b'_3-b'_2}{b_3-b_2} <0 \label{bc1}
\eeq
and
\beq
\frac{b'_3-b'_2}{b'_3-\tilde{b}'_1} =\frac{5}{12} \label{bc2}
\eeq
where the standard model renormalisation group coefficients are
\beq
b_3=-3 \ ,\ b_2=1 \ ,\ \tilde{b}_1=\mfrac{33}{5}
\eeq
and the $b'_i \ ,\ i=1,2,3$ which occur in the string loop threshold 
corrections \cite{deren,kaplun,dixon} are given by
\beqa
b'_3&=&9+2\sum^3_{g=1}(n_{Q(g)}+\frac{1}{2}n_{u(g)}+\frac{1}{2}n_{d(g)}) 
\label{b3}\\
b'_2&=&15+\sum^3_{g=1}(3n_{Q(g)}+n_{L(g)}) +n_{H_1} +n_{H_2} \label{b2}
\eeqa
and
\beq
\tilde{b}'_1=\frac{99}{5}+\frac{1}{5}\sum^3_{g=1}(n_{Q(g)} +8n_{u(g)}
+2n_{d(g)}+3n_{L(g)}+6n_{e(g)} ) +\frac{3}{5}(n_{H_1}+n_{H_2}) \label{b1}
\eeq
where the sum over $g$ is a sum over generations. Here the $U(1)$ coupling 
constant $\tilde{g}_1$ is normalised so that all three coupling constants are 
equal at the unification scale. Assuming generation universality to avoid 
flavour changing neutral currents,
\beq
n_{Q(1)}=n_{Q(2)}=n_{Q(3)}=n_Q
\eeq
and similarly for the modular weights of the other states, the solutions of 
(\ref{bc1}) and (\ref{bc2}) with modular weights given by (\ref{mw1}) and 
(\ref{mw2}) or (\ref{mw3}) and (\ref{mw4}) are given in table \ref{modws} with
\beq
M_{string} \approx 0.53\times g_{string} \times 10^{18} \mbox{GeV} \label{Mstr}
\eeq
and
\beq
g_{string}\approx 0.7\ .
\eeq
\begin{table}
\[
\begin{tabular}{ccccccc}
$n_{QL}$&$n_{UR}$&$n_{DR}$&$n_{LL}$&$n_{ER}$&$n_{H_1}$&$n_{H_2}$\\ 
\hline
0 & -2 & 1 & -3 & -1 & -1 & -1 \\
0 & -1 & 0 & -3 & -2 & -1 & -1 \\
0 & -1 & 1 & -2 & -2 & -1 & -1 \\
0 & -2 & 0 & -4 & -1 & -1 & -1 \\
0 & -1 & -1& -4 & -2 & -1 & -1 \\
\hline
\end{tabular}
\]
\caption{Modular weights \label{modws}} 
\end{table}
The corresponding value of $T$ for which unification takes place at
\beq
M_X \approx 2 \times 10^{16} \mbox{GeV} 
\eeq
is  given by \cite{deren,ibanez2,ibanez}
\beq
\frac{M_X}{M_{string}}=[(T+\bar{T})| \eta(T)|^4 ]^{(b'_3-b'_2)/2(b_3-b_2)}
\label{unif}
\eeq
and we find $T=14.5$ is suitable for all choices of modular weights of table 
\ref{modws}, and also gives the gauge couplings as $\alpha_s(m_Z)=0.115$ and 
$\sin^2\theta_W(m_Z)=0.2315$. 
We have restricted $n_{H_1}$ and $n_{H_2}$ to take 
the value 
$-1$ for consistency with the two mechanisms for generating the $\mu$ parameter
that we shall discuss in the next section, both of which require the Higgs 
fields to be untwisted sector states.

\section{Soft supersymmetry breaking terms at the string scale}
The soft supersymmetry scalar masses, gaugino masses and $A$ and $B$ terms
which occur in a supergravity theory may be calculated from the low energy 
limit of an orbifold compactification of the heterotic string given the 
K\"ahler potential, the superpotential and the gauge kinetic function derived
from the string theory \cite{cvetic,ibanez,carlos,brig,brig2}.In view of our 
current lack of detailed knowledge of
the non-perturbative superpotential responsible for supersymmetry breaking, 
a possible approach \cite{brig} is to absorb this uncertainty into an angle 
$\theta$
which measures the relative contributions of the dilaton and $T$ modulus 
auxiliary fields to supersymmetry breaking. In this section we summarize 
the resulting formulae \cite{brig} for the soft supersymmetry breaking terms 
and discuss
the choice of values for the Green-Schwarz parameters, the moduli dependence 
of the Yukawa couplings that occur in the $A$ term and the $\mu$ parameter that
occurs in the $B$ term.
The expressions for the soft supersymmetry breaking terms are expressed in 
terms of the angle $\theta$ defined by \cite{brig}
\beq
F^S-\dgst(T+\bar{T})^{-1}F^T=\sqrt{3}\,C\, Y m_{3/2} \sin{\theta} \label{FS}
\eeq
and
\beq
\left( \frac{Y-\frac{\dgst}{3}}{Y} \right)^{1/2} F^T=C\,(T+\bar{T})\,m_{3/2}
\cos{\theta}
\eeq
where the auxiliary fields $F^S$ and $F^T$ for the dilaton and the overall 
$T$ modulus are given in terms of $G \equiv K+\ln{|W|^2}$ by
\beq
F^S -\dgst(T+\bar{T})^{-1}F^T=Y^2 m_{3/2} \frac{\p G}{\p S}
\eeq
and
\beq
F^T=\frac{ (T+\bar{T})^2 Y m_{3/2}}{3\left(Y-\frac{\dgst}{3}\right)}
\left( \frac{\p G}{\p T}+\dgst(T+\bar{T})^{-1} \frac{\p G}{\p S}\right)\ \ .
\label{FT}
\eeq  
It has been assumed that all three moduli $T_i,\ i=1,2,3$, are on the same 
footing in the K\"ahler potential and superpotential and possible (CP-violating
) phases have been dropped for present purposes. The vacuum energy $V_0$ is 
given by
\beq
V_0=3(C^2 -1)m^2_{3/2}
\eeq
where 
\beq
m_{3/2}=e^{G/2} \label{grav}
\eeq
at the minimum of the effective potential.
Thus, if the vacuum energy is identified with the cosmological constant we
should take $C=1$. This we shall do throughout.

The soft supersymmetry breaking scalar masses, $m_\alpha$, are given by 
\beq
m^2_\alpha=(3C^2-2)m^2_{3/2}+n_\alpha C^2 m^2_{3/2} \frac{Y}
{\left(Y-\frac{\dgst}{3}\right)}\cos^2{\theta} \label{scalars}
\eeq
with overall modular weights $n_\alpha$ as in (\ref{n}). The gaugino masses 
$M_a$ are given by 
\beq
2m^{-1}_{3/2}(\mbox{Re}f_a) M_a= \sqrt{3} C Y \sin{\theta}+
\left(\frac{Y}{Y-\frac{\dgst}{3}}\right)^{1/2} C (T+\bar{T}) \cos{\theta}
\frac{(b'_a-\dgs)}{16 \pi^3} \,\hat{G}_2(T,\bar{T}) \label{gauginos}
\eeq
where
\beqa
\hat{G}_2(T,\bar{T})&=&G_2(T)-2\pi(T+\bar{T})^{-1} \\
G_2(T)&=&-4\pi \frac{d \ln{\eta}}{dT}
\eeqa
where $\eta(T)$ is the Dedekind function and, for the standard model, 
$b'_a, \ a=1,2,3$, are given by (\ref{b3})-(\ref{b1}). The real part of the 
gauge kinetic funcion Re$f_a$ is given by
\beq
\mbox{Re}f_a=g^{-2}_a(M_{string})
\eeq
and is determined from the gauge coupling constant $g_a(m_Z)$ at the 
electroweak scale $Q=m_Z$ by 
\beq
g^{-2}_a(M_{string})-g^{-2}_a(m_Z)=\frac{b_a}{16\pi^2} \ln \left(\frac{m_Z^2}
{M^2_{string}}\right)
\eeq
with $M_{string}$ as in (\ref{Mstr}).

The soft supersymmetry breaking $A$ terms $A_{\alpha\beta\gamma}$ are given by
\beq
A_{\alpha\beta\gamma}=-\sqrt{3} C m_{3/2} \sin\theta- Cm_{3/2}
\left(\frac{Y}{Y-\frac{\dgst}{3}}\right)^{1/2}\cos\theta\,\omega_{\alpha\beta
\gamma}(T)
\eeq
where
\beq
\omega_{\alpha\beta\gamma}(T)=3+n_\alpha+n_\beta+n_\gamma-(T+\bar{T})
\frac{\p \ln h_{\alpha\beta\gamma}}{\p T}
\eeq
and the trilinear term $\t{W}_3$ in the perturbative superpotential has 
been written as
\beq
\t{W}_3=h_{\alpha\beta\gamma} \phi_{\alpha} \phi_{\beta} \phi_{\gamma}\ \ .
\eeq
The modular weights $n_\alpha$, $n_\beta$ and $n_\gamma$ are chosen to 
correspond to one of the solutions for gauge coupling constant unification 
(for the $\zii$ or $\ziii$ orbifold) discussed in the 
previous section. Since we are assuming large values of Re$T$ in order to 
reduce the unification scale to $2 \times 10^{16}$ GeV, we shall use the 
asymptotic form of $h_{\alpha\beta\gamma}$ valid for large Re$T$,
\beq
h_{\alpha\beta\gamma} \sim \exp\left( - \frac{\pi}{3}\lambda_{\alpha\beta
\gamma} T \right)  \label{htop}
\eeq
where $\lambda_{\alpha\beta\gamma}$ is an integer in the range 0 to 4 for the
$\zii$ orbifold and in the range 0 to 10 for the $\ziii$ 
orbifold \cite{bl2}. The constant of proportionality in (\ref{htop}) is 
expected to be of order $g_{string}$.
Here and in (\ref{scalars}) and (\ref{gauginos}) we shall
use 
the Green-Schwarz parameter obtained by inserting $\dgs^i,\ i=1,2,3$, for 
the $\zii$ or $\ziii$ orbifold in (\ref{gs}) and 
(\ref{gst}).
Although, in general, the Green-Schwarz parameters $\dgs^i$ have different 
values for the different complex planes, contradicting the assumption that 
all three complex planes are on the same footing in $G$, this better 
approximates the situation than neglecting the Green-Schwarz parameters. A
simple model is to take a pure gauge hidden sector with $E_8$ gauge group.
Then \cite{deren,bl3} 
\beq
\dgs^i=\frac{b_a}{3} \left(1-2\frac{|G_i|}{|G|}\right)
\eeq
where $a$ now refers to the hidden sector gauge group and the $i$th complex
plane is left unrotated by the subgroup $G_i$ of the point group $G$. With 
$b_a=-90$ for $E_8$ we have 
\beq 
\dgs=-40\mbox{ or }-50 \label{gsvals}
\eeq
for $\zii$ or $\ziii$ respectively.

The soft supersymmetry breaking $B$ term is more model dependent because of 
different possible origins for the $\mu$ parameter. If the $\mu$ term is 
generated non-perturbatively as an explict superpotential term $\mu_W \phi_1
\phi_2$, where $\phi_1$ and $\phi_2$ are the superfields for the Higgs scalars
$H_1$ and $H_2$, then the $B$ term, which in this case we denote by $B_W$, is 
given by
\beqa
m^{-1}_{3/2}B_W&=&-1-\sqrt{3}C\sin\theta\left(1-Y\frac{\p\ln\mu_W}{\p S}\right)
-\left(\frac{Y}{Y-\frac{\dgst}{3}}\right)^{1/2} C \cos\theta\left(\frac{}{}3+
n_{H_1}+n_{H_2}\right.\nonumber\\
& &\ \ \ \ \ \left.-(T+\bar{T})\frac{\p\ln\mu_W}{\p T}-\dgst\frac{\p\ln\mu_W}
{\p S} \right) \ \ . \label{BW} 
\eeqa
If the $\mu$
parameter is gaugino condensate induced \cite{antoniadis} then 
\beq
\mu_W \propto W_{np}\frac{\p\ln\eta(T_3)}{\p T_3}\frac{\p \ln\eta(U_3)}{\p 
U_3} \label{muW}
\eeq
where $W_{np}$ is the non-perturbative superpotential and the orbifold is 
assumed to possess a $Z_2$ plane, taken to be the third complex plane with 
associated moduli $T_3$ and $U_3$ and untwisted matter fields $\phi_1$ and
$\phi_2$. Such a mechanism is possible for the $\zii$ orbifold though not for 
the $\ziii$ orbifold which does not possess a $Z_2$ plane. In the case of 
$\ziii$ we take $\mu_W$ constant as in ref.\cite{brig}.
Because $\phi_1$ and
$\phi_2$ are then necessarily untwisted states the modular weights $n_{H_1}$ 
and
$n_{H_2}$ should be taken to be $-1$. It is somewhat problematic to employ 
this mechanism in the context of the simple model with only a single overall 
modulus $T$ being considered here. However if we neglect the auxiliary field 
for $U_3$, or equivalently assume that $U_3$ does not contribute 
significantly to the supersymmetry breaking, then (\ref{BW}) is correct when 
$T_1$, $T_2$ and $T_3$ are on the same footing. There is also the difficulty
that gaugino condensate models in general produce a negative vacuum energy
$V_0$ rather than zero vacuum energy, as we have assumed after (\ref{grav}).
Nonetheless, we think it worthwhile to study this mechanism to obtain a 
flavour of the effect on the supersymmetric particle spectrum of the kind of
moduli dependence of the $\mu$ parameter that can occur in physically 
motivated models. After evaluating $\frac{\p\ln\mu_W}{\p S}$ and 
$\frac{\p\ln\mu_W}{\p T}$ we obtain
\beq
m^{-1}_{3/2}B_W=3C^2-1-\left(\frac{Y}{Y-\frac{\dgst}{3}}\right)^{1/2} C \cos 
\theta \left(n_{H_1}+n_{H_2}-(T+\bar{T}) \left( \frac{\p\ln\eta(T)}{\p T}
\right)^{-1}\frac{\p^2\ln\eta(T)}{\p T^2} \right)
\eeq
Here $\frac{\p\ln\mu_W}{\p S}$ and $\frac{\p\ln\mu_W}{\p T}$ have been written
in terms of $F^T$ and $F^S$ and so in terms of $\theta$ using 
(\ref{FS})-(\ref{FT})
and because $T_i,\ i=1,2,3$, are not on the same footing in 
(\ref{muW}), $(T+\bar{T})\frac{\p\ln\mu_W}{\p T}$ has been interpreted as 
$(T_3+\bar{T}_3)\frac{\p\ln\mu_W}{\p T_3}$ evaluated at $T_3=T$. 

On the other hand if the $\mu$ parameter is generated by a term of the form 
$(Z \phi_1 \phi_2+h.c.)$ in the K\"ahler potential \cite{antoniadis} mixing 
the Higgs 
superfields then the tree level form of $Z$ is
\beq
Z=(T_3+\bar{T}_3)^{-1} (U_3+\bar{U}_3)^{-1}
\eeq
if the third complex plane is the $Z_2$ plane with whose moduli, $T_3$ and 
$U_3$, the untwisted matter fields $\phi_1$ and $\phi_2$ are associated for
this mechanism. The effective $\mu$ parameter $(\mu_Z)_{eff}$ derived from 
the Higgsino mass term is 
\beq
(\mu_Z)_{eff}=|W_{np}|Z(1+C\cos\theta) \label{muZ}
\eeq
and the final form of the $B$ term, which in this case we denote by $B_Z$ is 
given by
\beq
m^{-1}_{3/2} B_Z= \frac{[2(1+C\cos\theta)-3(C^2-1)]}{(1+C \cos\theta)}\ \ .
\label{BZ}
\eeq
In particular, when $V_0$ is zero so that $C=1$, as we are assuming throughout
, $B_Z$ takes the constant value $2m_{3/2}$. This compares with 
$2(1+\cos\theta)m_{3/2}$ in ref. \cite{brig}, where $Z$ was taken to be a 
moduli 
independent constant. In arriving at (\ref{muZ}) and (\ref{BZ}) we have again 
assumed that there is no significant supersymmetry breaking due to the $U$ 
modulus, so as to be able to neglect the auxiliary field for the $U$ modulus. 
Also, here and elsewhere in this section the usual rescaling by a factor 
$e^{K/2}\frac{\bar{W}_{np}}{|W_{np}|}$ required to go from the supergravity 
theory derived from the orbifold compactification of the string theory to the 
globally supersymmetric theory has been carried out, together with 
normalisation of the matter fields (see, for example, ref.\cite{brig}).

\section{Running of coupling constants and supersymmetry breaking parameters}
The method for running coupling constants and supersymmetry breaking parameters
from the unification scale $M_X$ to the electroweak scale is well known. (See,
for example, refs. \cite{ibanez3} and \cite{ibanez4}.) The relevant 
renormalisation group equations
for our purposes are summarized in appendix A  and the relevant solutions in 
appendix B, with the bottom quark and $\tau$ lepton Yukawa couplings, as well 
as the first and second generation Yukawa couplings, neglected but the effect 
of the $\mu$ parameter retained. The top Yukawa, $h_t$, and the $\mu$ parameter
have been defined  through the superpotetial terms
\beq
W=h_tQ_tt^cH_2-\mu H_1H_2 \label{W}
\eeq
where $Q_t$ is the doublet $(t\ b)_L$ 
for the top and bottom quarks, $t^c$ is the corresponding singlet and the 
Higgs doublets are
\beq
H_1=\left(\begin{array}{c} H^0_1 \\H^-_1 \end{array}\right) \ ,\
H_2=\left(\begin{array}{c} H^+_2 \\H^0_2 \end{array}\right)
\eeq
where $H^0_1$ and $H^0_2$ have expectation values $v_1$ and $v_2$ respectively.
In (\ref{W}), $Q_t t^c H_2$ is shorthand for $Q_t^T i \tau^2 H_2 t^c$ and
$\mu H_1 H_2$ for $\mu H^T_1 i \tau^2 H_2$.
The tree level Higgs scalar potential $V_{eff}$ in terms of the above 
expectaton values is 
\beq
V_{eff}=\mu^2_1 v^2_1+\mu^2_2 v^2_2-2\mu^2_3v_1v_2+\frac{1}{8}(g^2_1+g^2_2)
(v^2_1-v^2_2)^2
\eeq
where
\beq
\mu^2_1=m^2_1+\mu^2 \ ,\ \mu^2_2=m^2_2+\mu^2 \ ,\ \mu^2_3=-\mu B m_{3/2}
\eeq
and $m_1$ and $m_2$ are the soft supersymmetry breaking masses for $H_1$ and 
$H_2$. Minimisation of the tree level effective potential gives 
\beq
\omega^2 = \frac{\mu_1^2+\frac{1}{2}m_Z^2}{\mu_2^2+\frac{1}{2}m_Z^2}
\label{min1} 
\eeq
and
\beq
\frac{\omega}{\omega^2+1}=\frac{\mu_3^2}{\mu_1^2+\mu_2^2} \label{min2}
\eeq  
at the electroweak scale $Q=m_Z$ where
\beq
\omega^{-1}=\tan\t{\theta}=\frac{v_1}{v_2}\ \ .
\eeq
Also the following inequalities must hold
\beqa
\mu_1^2+\mu_2^2&>&2|\mu_3^2| \label{con1}\\
\mu_3^4&>&\mu_1^2\mu_2^2  \label{con2}\\
\mu_2^2+m_{QL}^2+m_{UR}^2&>&m^2(2|A_t|-3)\label{ccb}
\eeqa
as explained in ref.\cite{ibanez3}.

Our strategy for fixing some of the parameters in the models is as follows. 
Knowing the values $m_1^2(0)$, $m_2^2(0)$ and $B(0)$ of the soft supersymmetry
breaking parameters at the gauge coupling constant unification scale $M_X$ 
(which differ little from their values at the string scale) and assuming values
for the gravitino mass $m_{3/2}$ in the range 100 GeV to 10 TeV, (\ref{min1})
and (\ref{min2}) are a pair of equations that can be solved for $\mu(0)$ and
$\omega$. Then
\beq
m_W^2=\frac{g_2^2}{2}(v_1^2+v_2^2)
\eeq
determines $v_1$ and $v_2$. In addition 
\beq
m_t=h_t v_2
\eeq
fixes $h_t$ at the electroweak scale and eqn.(\ref{Yt}) determines $h_t (0)$.
We run all renormalisation group equations from the gauge coupling constant 
unification scale $M_X$ and ignore small effects due to the difference between
$M_X$ and the string scale.

The supersymmetry breaking parameters at the string scale are calculated as in
$\S 3$. Then the predictions for the supersymmetric particle spectrum to be
discussed in the next section are parameterised by the angle $\theta$ which 
measures the relative contribution of the dilaton $S$ and the modulus $T$ to 
supersymmetry breaking and the gravitino mass (assuming zero vacuum energy
$V_0$ so that $C=1$). In addition, the outcome for the spectrum depends on 
the choice of modular weights $n_\alpha$ from amongst the sets allowed for 
the $\zii$ and $\ziii$ orbifolds, as in table \ref{modws}, and on the 
mechanism 
adopted to generate the $\mu$ parameter, which influences the form of the $B$ 
term. The choice of modular weights also fixes the value of $T$ from 
(\ref{unif}). 
The Green-Schwarz parameters $\dgs$ are taken from (\ref{gsvals}).

The above discussion neglects radiative corrections to the effective potential.
When these are included \cite{ellis,ellis2} the strategy for obtaining the 
expectation values 
$v_1$ and $v_2$ has to be amended. Those supersymmetric particle masses that 
depend on $v_1$ and $v_2$ are then modified as well as the Higgs scalar masses.
We will discuss these points in detail in the next section. We have not 
considered the radiative corrections to (\ref{con1})-(\ref{ccb}) which may 
exclude some values of $\theta$, in particular the dilaton dominated case 
\cite{casas}.

\section{The supersymmetric particle spectrum}
The expressions for the masses of the supersymmetric partners of standard model
states in terms of the soft supersymmetry breaking parameters are well known.
For the first two generations of quarks and leptons the Yukawa couplings and 
$A$ terms are negligible and the corresponding squark and slepton mass terms
are simply the soft supersymmetry breaking scalar masses. For the third 
generation it is necessary to allow for a non-negligible top Yukawa coupling 
and the top squark masses are given by
\beq
m^2_{\t{t}_{h,l}}=m_t^2+\frac{1}{2}\left(m^2_Q+m^2_U \pm \left[(m^2_Q-
m^2_U)^2+4m_t^2(A_tm_{3/2}+\mu\omega^{-1})\right]^{1/2}\right)
\eeq
where $m_Q$ and $m_U$ refer to the scalar partners of the quark doublet and 
one of the quark singlets for the third generation respectively, and the 
D term has been neglected.

The gluino mass is given by the Majorana mass term. However, the Wino and Zino
mix with the Higgsinos. The chargino mass matrix has eigenvalues $m_{c_{h,l}}$ 
given by
\beq
2m^2_{c_{h,l}}=M_2^2+\mu^2+2m_W^2 \pm \Delta^{1/2}
\eeq
where
\beq
\Delta = (M_2^2-\mu^2)^2 +4m_W^2(M_2^2 +\mu^2 +2 M_2 \mu \sin{2\t{\theta}})
+4m_W^4 \cos^2 2\t{\theta}
\eeq
The neutralino mass matrix has the form
\beq 
\bmat{cc}
 &\bmat{cccc}\ i\t{W}^3\ \ &\ i\t{B}\ \ &\ \ \t{h}_2^0\ &\ \ \t{h}_1^0\ \emat \\
\bmat{c}i\t{W}^3\\i\t{B}\\\t{h}_2^0\\\t{h}_1^0 \emat 
& \left(\bmat{cccc}-M_2&0&-\frac{g_2v_2}{\sqrt{2}}&\frac{g_2 v_1}{\sqrt{2}}\\
             0&-M_1&\frac{g_1v_2}{\sqrt{2}}&-\frac{g_1v_1}{\sqrt{2}}\\
             -\frac{g_2v_2}{\sqrt{2}}&\frac{g_1v_2}{\sqrt{2}}&0&\mu\\
            \frac{g_2 v_1}{\sqrt{2}}&-\frac{g_1v_1}{\sqrt{2}}&\mu&0\emat\right)
\emat \ \ +h.c.
\eeq
In addition the charged Higgs has mass
\beq
m_{H^\pm}^2 =m_W^2 +\mu_1^2 +\mu_2^2
\eeq
and the neutral Higgses have masses
\beq
m_c^2=\mu_1^2+\mu_2^2
\eeq
and
\beq
m_{a,b}^2=\frac{1}{2}\left(m_c^2+m_Z^2\pm\left[(m_c^2+m_Z^2)^2-4m_c^2m_Z^2
\cos^2 2\t{\theta} \right]^{1/2}\right) \ . \label{neutralhiggs}
\eeq

In our detailed calculations, the mass $m_b$ given by (\ref{neutralhiggs}) is 
generically lower than the experimental bound. However, the one loop radiative
corrections to the Higgs scalar masses are substantial \cite{ellis,ellis2} and
we shall use the 
one loop Higgs scalar effective potential in what follows. The one loop 
corrected formulae for the Higgs masses can be found in ref.\cite{ellis2}. 
When the one
loop corrections to the effective potential are included the minimisation 
conditions (\ref{min1}) and (\ref{min2}) for the expectation values $v_1$ and
$v_2$ are modified with the result that
\beq
\omega^2= \frac{2\mu_1^2+M_Z^2+\frac{1}{v_1}\frac{
\p\Delta V_1}{\p v_1}-\frac{v_2}{v_1^2}\frac{
\p\Delta V_1}{\p v_2}  }{2\mu_2^2+m_Z^2} \label{1lpmin1}  
\eeq
and
\beq
\frac{\omega}{\omega^2+1}=\frac{2\mu_3^2}{2\mu_2^2 +2\mu_1^2 + 
\frac{1}{v_1}\frac{\p \Delta V_1}{\p v_1}+\frac{1}{v_2}\frac{
\p \Delta V_1}{\p v_2}} \label{1lpmin2}
\eeq 
where $\Delta V_1$ is the one loop correction to the effective potential 
evaluated at $m_Z$ and 
\beqa
\frac{\partial\Delta V_1}{\partial v_i}&=&\frac{3}{16\pi^2}\left[
m_{\tilde{t}_h}^2 \frac{\partial m_{\tilde{t}_h}^2}{\partial v_i} \left(
\ln{\frac{m_{\tilde{t}_h}^2}{m_Z^2}}-1 \right) +
m_{\tilde{t}_l}^2 \frac{\partial m_{\tilde{t}_l}^2}{\partial v_i} \left(
\ln{\frac{m_{\tilde{t}_l}^2}{m_Z^2}}-1 \right) \right. \nonumber\\
& &\left.-2m_t^2 \frac{\partial m_t^2}{\partial v_i} 
\left(\ln{\frac{m_t^2}{m_Z^2}}-1\right)\right] \ ,\ i=1,2\ \ .
\eeqa

The strategy for fixing the parameters in the models is essentially that 
described in $\S 4$ except that (\ref{1lpmin1}) and (\ref{1lpmin2}) should now
be 
regarded as a pair of equations for $v_1$, $v_2$ and $\mu(0)$ given the soft 
supersymmetry breaking parameters $m_Q^2$, $m_U^2$ and $A_t$ at the string 
scale and given values for $m_t$ and $m_{3/2}$, rather than as a pair of 
equatons that can be solved for $\mu(0)$ and $\omega$.

In deriving the possible supersymmetric particle spectrum we have insisted on 
no negative squared masses at the string scale to avoid high scale symmetry 
breaking in the standard model. We have also insisted on the following 
experimental constraints.  From LEP1.5 data, there are no charged or coloured 
sparticles with 
masses less than 65 GeV, the lightest Higgs is heavier than 65 GeV and the 
lower bound on the charginos is 80~GeV. Tevatron data indicates that the 
gluino mass is above 175 GeV, but should not exceed 1.5 TeV (to avoid 
reintroducing the hierarchy problem). The top quark mass is known to be 
175$\pm$6 GeV. The vev of the Higgs responsible for the top quark mass has a
maximum value given by 
\beq
v_1^2+v_2^2 = \frac{2 m_Z^2}{(g^2+g'^2)}
\eeq
with $v_1^2=0$ implying $v_2$(max)=173.3 GeV.
Since $m_t=h_t v_2$ this puts a lower limit on $h_t$ if $m_t=175 \pm6$
GeV is to be obtained. Specifically the value at $M_X$ is $h_t(min)=0.52 $ and 
so it is appropriate to set $\lambda=0$ in (\ref{htop}) for the top Yukawa 
coupling. One loop minimisation conditions have been used throughout
 and the Higgs masses are one loop corrected. The parameter $\omega$ is found 
to be never greater than 6, justifying the neglect of the b-quark contibution.
The D terms have been included in the mass of the lightest sleptons, the right
selectron and the left sneutrino. In the figures the following notation is 
used for the masses:
\newline
$c_h$ , $c_l$ : heavy and light charginos
\newline $t_h$ , $t_l$ : heavy and light stops
\newline  $H_a$ , $H_b$ : heavy and light CP-even Higgses respectively
\newline  $m_t$ : top quark 
\newline  $E_R$ , $V_L$ : right selectron and left sneutrino respectively
\newline  $N1$ : lightest neutralino
\newline $g$ : gluino \newline
Particles not displayed are the three neutralinos which are degenerate with 
the charginos, the charged and CP-odd Higgses which are degenerate with $H_a$,
the remaining squarks which are all only slightly less massive than the gluino,
and the left selectron which is always heavier than $V_L$. 

Several models will now be presented that are representative of the variety of
the supersymmetric particle spectra that can occur.
\begin{figure}[p]
\[
\epsfxsize=6in
\epsfysize=3.5in
\epsffile{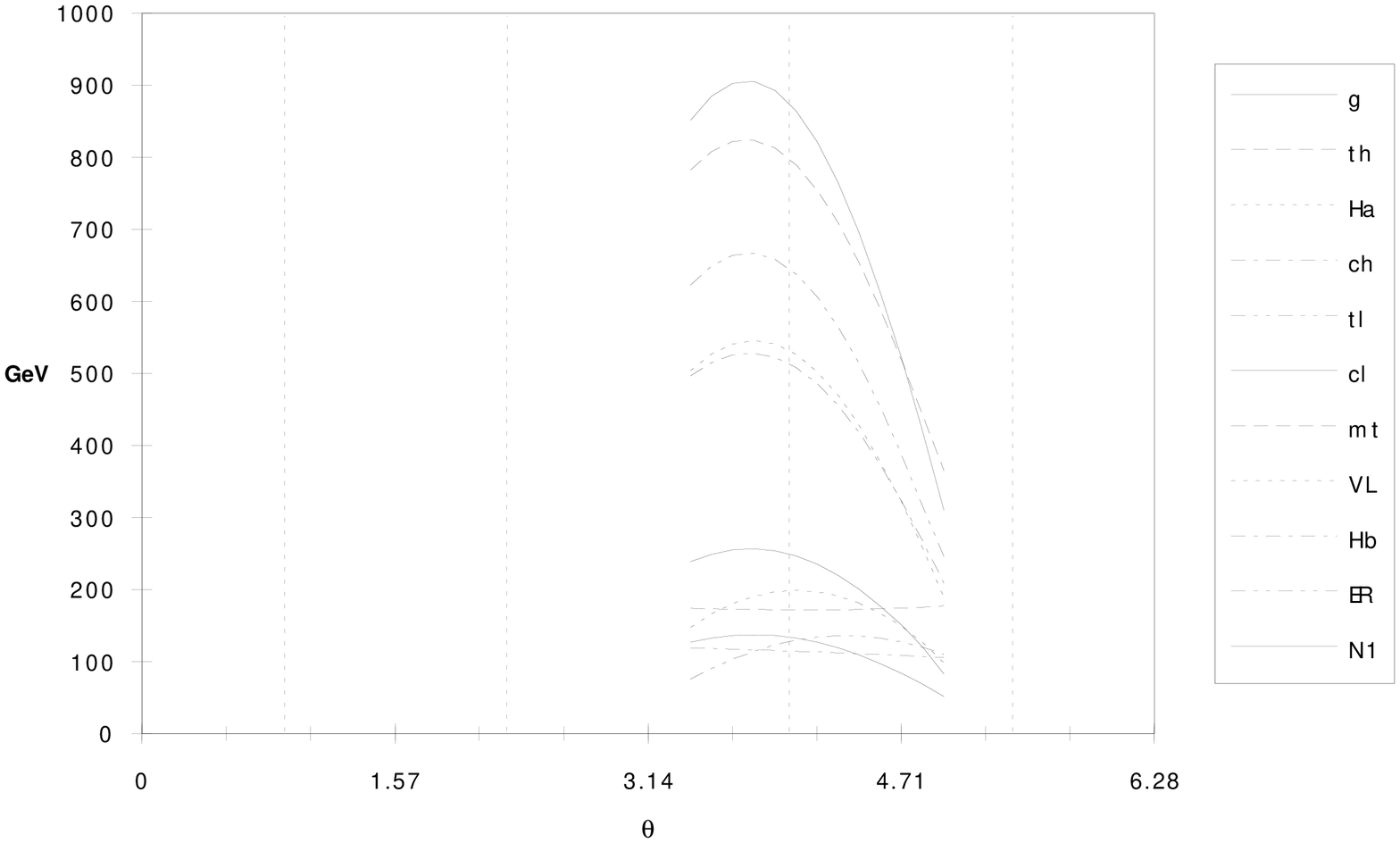}
\]
\caption{$\dgs=-50$, $m_{3/2}=100$ GeV, $T=14.5$, $h_t=0.7$,
$B \equiv B_W$, modular weights as in line 2 of table \ref{modws}.
\label{modA4}}
\end{figure}
\begin{figure}[p]
\[
\epsfxsize=6in
\epsfysize=3.5in
\epsffile{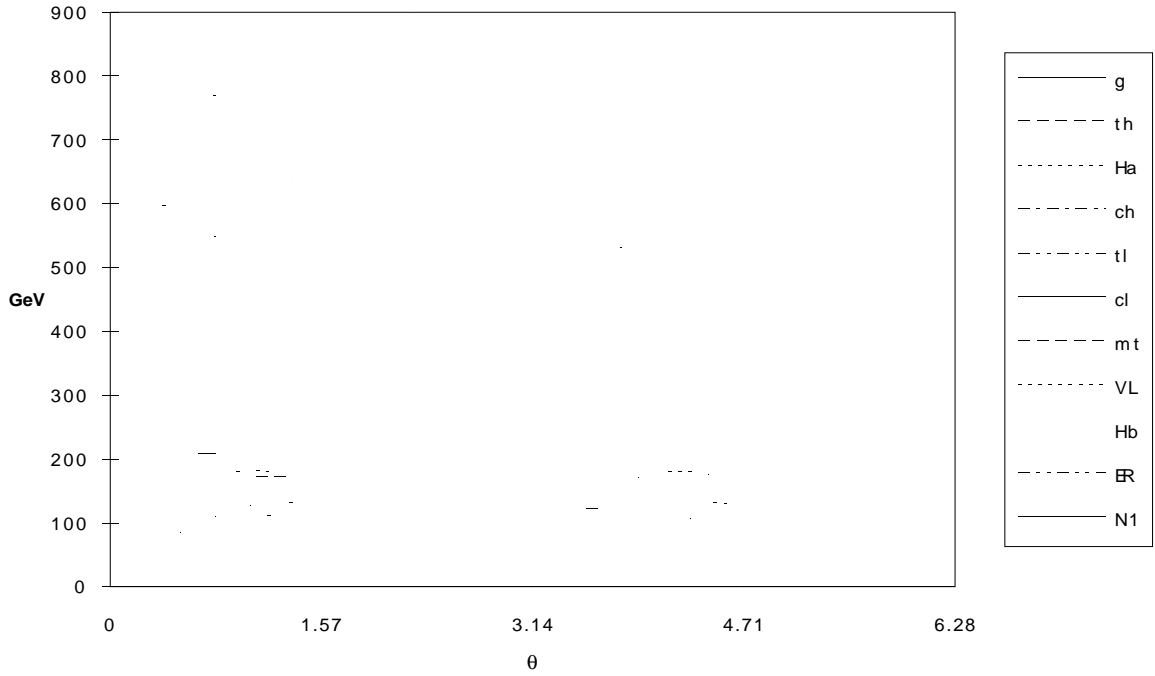}
\]
\caption{$\dgs=-40$, $m_{3/2}=100$ GeV, $T=14.5$, $h_t=0.9$, $B \equiv B_W$, 
modular weights as in line 2 of table \ref{modws} \label{modB2}}
\end{figure}

In figure \ref{modA4} the resulting mass spectrum is shown for the $\ziii$ 
orbifold, characterised here by $\dgs=-50$, with $m_{3/2}=100$ GeV, $h_t=0.7$, 
modular 
weights as in line 2 of table \ref{modws} and with the $B$ term given by $B_W$
with $\mu_W$ constant.
The only allowed region is bounded on the left by $E_R$ acquiring a 
too low mass, and on the right by $c_l$ becoming too light while $m_t$ is 
always in the vicinity of 175 GeV. 
Further, the acceptable part of the spectrum is limited by the
requirement of positive squared scalar masses at the string scale which 
confines it to the regions between the two pairs of vertical lines shown in 
the figure, centred on the dilaton dominated limits ($\theta=\frac{\pi}{2},
\frac{3\pi}{2}$). The part of the spectrum around $\theta=\frac{\pi}{2}$ is 
ruled out due to $m_t$ being unacceptably low and the electroweak symmetry 
is unbroken there.

As in ref.\cite{brig}, there is a clear division of the 
spectrum into a heavy and a light group although now
there is a far greater variation in masses as $\theta$ is varied than was the 
case in ref.\cite{brig}.  This latter effect is attributable directly to the 
magnitude 
of $\dgs$ and the effect it has on the gaugino masses which feed through to 
all sparticles.  At the dilaton dominated limit the spectrum is 
qualitatively similar to that in ref.\cite{brig} but away from this limit we 
see that the gluino (and squarks) are often heavier than the heavy stop 
(in \cite{brig} the gluino mass was fixed).
Particularly noticeable, and attributable to $|\dgs|$, is the mass  
of the lightest neutralino (which is mostly $M_1$) which often exceeds 100 
GeV. This is worth 
noting because, as seen in figure \ref{modA4}, on the left hand limit of the 
allowed region its role 
as the `lightest supersymmetric particle' is jeopardised in favour of the 
right selectron.
This is why the D term has been included in $E_R$
(it can add 10 GeV or more). The lightest Higgs, $H_b$, is also in the region 
of 100 GeV and the light chargino can be lighter than the sleptons.

A change in $m_{3/2}$ will scale all masses (except $m_t$). Decreasing 
$m_{3/2}$
narrows the allowed region by virtue of $E_R$ and $c_l$ 
becoming too light on the right hand edge, giving an 
approximate effective minimum of $m_{3/2}\simeq70$ GeV, below which the 
dilaton dominated limit is unreachable. Increasing $m_{3/2}$ 
rapidly increases the gluino mass to way above the limit 1.5 TeV. For 
$m_{3/2}\geq250$ GeV even the dilaton dominated limit is excluded. 
The allowed regions may not be extended to the right 
significantly, even with a high gravitino mass  because $c_l$ remains too light
there. A variation in   
$h_t$ affects all masses due to its appearance in the one-loop effects and an 
increase in $h_t$ will increase all masses slightly. However adjustment of 
$h_t$ is restricted by $m_t$ and it may not deviate far from $0.7$ without 
pushing $m_t$ out of the experimental bounds. 

A different choice of modular weights from table \ref{modws} will not change 
the spectrum very much other than by changing the acceptable width at the 
string scale. The modular weights from line 3 of table \ref{modws} are the 
least restrictive
at the string scale, the `heaviest' weight being $-2$, and so the acceptable 
part of the spectrum is widened slightly, while lines 4 and 5 from table 
\ref{modws} have the opposite effect. 
Thus the spectrum displayed in figure \ref{modA4} is very typical and 
deviations from it are small.


\begin{figure}[tb]
\[
\epsfxsize=6in
\epsfysize=3.5in
\epsffile{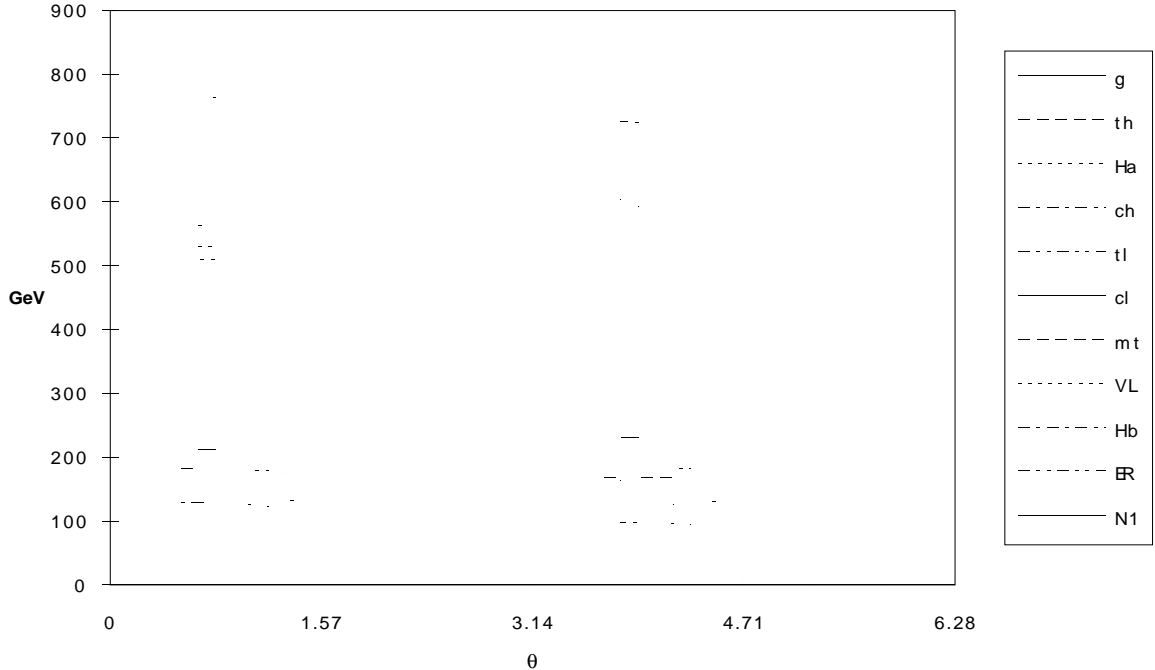}
\]
\caption{$\dgs=-40$, $m_{3/2}=100$ GeV, $T=14.5$, $h_t=0.9$, $B \equiv B_Z$, 
modular weights as in line 2 of table \ref{modws}. \label{modJ2}}
\end{figure}
Figure \ref{modB2} shows that spectrum obtained for a $\zii$ orbifold 
($\dgs=-40$) with 
$\mu_W$ as in (\ref{muW}) has two valid regions. The inclusion of the 
derivatives of $\mu_W$ in $B_W$ is instrumental in obtaining this result, were
they to be neglected we would obtain only one valid region similar to figure
\ref{modA4}.
Each region is qualitatively
similar to that shown in figure \ref{modA4}, although both regions are bounded
on the right by $m_t$ becoming too low. 
The dilaton dominated 
limit is not reachable at $\theta=\frac{3\pi}{2}$ for this reason, 
while that at $\theta=\frac{\pi}{2}$  is reachable. Note that 
here $h_t$ cannot deviate far from 0.9 without $m_t$ being pushed outside the 
experimental bounds. Conversely, obtaining an acceptable $m_t$ beyond the 
displayed regions 
would require an unacceptably high value of $h_t$.
We also find $70<m_{3/2}<250$ GeV for an acceptable spectrum, as 
before.  

Concerning the other form of the $B$ term, $B_Z$, which is only valid for the 
\mbox{$\zii$} orbifold because it requires a $Z_2$ plane, an example is 
shown in figure 
\ref{modJ2}. Comparison with figure \ref{modB2} shows some differences. In the 
right hand region the $H_a$-$H_b$ splitting is increased and in both regions  
the
$t_h$-$t_l$ splitting is reduced. There is near degeneracy between $c_h$, 
$t_l$ and $H_a$ in the right hand region. The light group remains relatively 
unaffected, although the top quark is, on average, heavier in the 
left hand region than the right hand region. To obtain central values of $m_t$ 
the left and right regions require $h_t=0.8, 1.0$ respectively. The dilaton 
dominated limits at $\theta=\frac{\pi}{2},\frac{3\pi}{2}$ are respectively 
included and excluded as in figure \ref{modB2}.

\section{Conclusions}
We have studied the supersymmetric particle spectrum in orbifold 
compactifications of string theory where unification of gauge coupling 
constants at $2\times 10^{16}$ GeV is due to large moduli dependent string 
loop threshold corrections, making a number of changes to the assumptions in 
Brignole {\em et al}.\ \cite{brig} in order to explore the robustness of the 
qualitative features of the spectrum obtained. 
The specific orbifold models considered here show that the inclusion of the 
derivatives of $\mu_W$ in $B_W$ is important in obtaining acceptable 
spectra in two separate regions and that it is also necessary for 
$h_t\approx0.9$ to obtain correct values for the top quark mass. For $\zii$ 
orbifold models there is then little resultant difference between the $B_Z$ and
$B_W$ mechanisms.    
For both mechanisms the 
gravitino mass is restricted approximately to the range 70-250 GeV in order 
to satisfy the upper and lower mass limits imposed in $\S5$. 
It is also 
apparent that the effects of the Green-Schwarz coefficient are not negligible.
In the 
examples presented here, $|\dgs|$ is substantial enough to shift the 
spectrum away from the dilaton dominated limit and partly out of the regions 
allowed at the string scale resulting in a considerably narrower acceptable 
range for $\theta$.
The lightest neutralino is often heavier than usually 
assumed ($\sim$100 GeV as opposed to $\sim$50 GeV \cite{brig}), and can be of 
similar mass to (or heavier than) the light Higgs and the right selectron.
In addition $|\dgs|$ induces a large variation in 
the masses as $\theta$ is varied, particularly for the heavy group.  
In principle this should make the goldstino angle, $\theta$, easier to 
determine if sparticles are eventually discovered.

\section*{Acknowledgments}
We are grateful to George Kraniotis for helpful discussions. This research was
supported in part by PPARC and P.S. was supported by a Royal Holloway 
studentship.

\appendix
\section{Renormalisation group equations}
The one loop renormalisation group equations for the various coupling 
constants and soft supersymmetry breaking parameters defined in the text, 
including the contribution of the $\mu$ parameter are as follows. In all cases 
the bottom quark and $\tau$ lepton Yukawa coupling have been neglected. 

The gauge coupling constants $g_i,\ i=1,2,3$, obey
\beq
\frac{d g_i^2}{d\ln Q}=\frac{b_i}{8\pi^2}g_i^4 \ ,\ i=1,2,3
\eeq
with normalisation of the $U(1)$ coupling constant such that
\beq
g_3^2(M_X)=g_2^2(m_X)=\frac{5}{3}g_1^2(M_X)
\eeq
at the unification scale $M_X$, then
\beq
b_3=-3,\ b_2=1,\ b_1=11
\eeq
and the corresponding gaugino masses obey
\beq
\frac{d M_i}{d\ln Q}=\frac{b_i}{8 \pi^2}g_i^2 M_i\ \ .
\eeq
The renormalisation group equation for the top quark Yukawa coupling $h_t$ is
\beq
\frac{d Y_t}{d t}=\frac{d \ln E}{dt}Y_t-6Y_t^2
\eeq
where
\beqa
t &\equiv& \ln\left(\frac{M_X^2}{Q^2}\right) \\
Y_t &\equiv& \frac{h_t^2}{16 \pi^2}
\eeqa
and
\beq
E(t)=(1+\beta_3 t)^{\frac{16}{3b_3}} (1+\beta_2 t)^{\frac{3}{b_2}} 
(1+\beta_1t)^{\frac{13}{9 b_1}}
\eeq
with
\beq
\beta_i=b_i\t{\alpha}_i(0)
\eeq
and
\beq
\t{\alpha}_i(t)=\frac{\alpha_i (t)}{4 \pi}=\frac{g_i^2 (t)}{16\pi^2}\ \ .
\eeq
The $\mu$ parameter obeys
\beq
\frac{d\mu^2}{dt}=(3\t{\alpha}_2 +\t{\alpha}_1-3Y_t)\mu^2\ \ .
\eeq
The soft supersymmetry breaking $A_t$ and $B$ parameters obey
\beq
\frac{dA_t}{dt}=m^{-1}_{3/2}\left(\frac{16}{3} \t{\alpha}_3M_3+3\t{\alpha}_2
M_2+\frac{13}{9}\t{\alpha}_1M_1\right)-6 Y_t A_t \label{A}
\eeq
\beq
\frac{dB}{dt}=m^{-1}_{3/2}(3 \t{\alpha}_2M_2+ \t{\alpha}_1M_1)-3Y_tA_t\ \ .
\eeq
The corresponding equations for all other $A$ terms are obtained by deleting 
the $Y_t$ term in (\ref{A}).
Scalars $\phi_\alpha$ that are the supersymmetric partners of the first and
second generation quarks and leptons have soft supersymmetry breaking masses
$m_\alpha$ obeying
\beq
\frac{dm_\alpha^2}{dt}=4 \sum_{i=1}^{3}C^\alpha_i \t{\alpha}_iM_i^2
\eeq
where the group theory factors $C^\alpha_i$ have the values $C^\alpha_3=\frac
{4}{3}$ for an $SU(3)_C$ triplet, $C^\alpha_2=\frac{3}{4}$ for an $SU(2)_L$
doublet, and $C^\alpha_1=Y^2$ for a state with weak hypercharge $Y$.

The renormalisation group equations for the masses $\mu_1$ and $\mu_3$ in the
Higgs scalar potential are 
\beq
\frac{d \mu_1^2}{dt}=(3\t{\alpha}_2M_2^2 +\t{\alpha}_1M_1^2) +
(3\t{\alpha}_2 +\t{\alpha}_1-3Y_t)\mu^2
\eeq
and
\beq
\frac{d \mu_3^2}{dt}=\left(\frac{3}{2} \t{\alpha}_2+\frac{1}{2}\t{\alpha}_1-
\frac{3}{2}Y_t\right)\mu_3^2 +3\mu m_{3/2} Y_t A_t-
\mu(3\t{\alpha}_2M_2+\t{\alpha}_1M_1)\ \ .
\eeq

The renormalisation group equations for the masses of the scalars which are 
the supersymmetric partners of the third generation quarks and leptons are 
expressed conveniently \cite{ibanez3} in terms of the quantities 
\beqa
m_4^2&=&m_D^2+m_U^2-2m_Q^2 \nonumber\\
m_5^2&=&\frac{2}{3}(\mu_2^2-\mu^2)-m_U^2 \nonumber\\
m_6^2&=&\frac{3}{2}m_D^2+m_L^2-(\mu_1^2-\mu^2) \nonumber\\
m_7^2&=&m_L^2-\frac{1}{2}m_E^2
\eeqa
where $m_Q$ and $m_L$ refer to the scalar partners of the quark and lepton
doublets and $m_U$, $m_D$ and $m_E$ refer to the scalar partners of the quark 
and lepton singlets for the third generation. Then the remaining 
renormalisation group equations for these masses and for the mass $\mu_2$ in 
the Higgs scalar potential are
\beqa
\frac{dm_4^2}{dt}&=&-6\t{\alpha}_2M_2^2+2\t{\alpha}_1M_1^2 \nonumber\\
\frac{dm_5^2}{dt}&=&-\frac{16}{3}\t{\alpha}_3 M_3^2 + 2 \t{\alpha}_2 M_2^2 -
\frac{10}{9}\t{\alpha}_1 M_1^2 \nonumber\\
\frac{dm_6^2}{dt}&=&8\t{\alpha}_3 M_3^2 +\frac{2}{3}\t{\alpha}_1 M_1^2 
\nonumber\\
\frac{dm_7^2}{dt}&=&3\t{\alpha}_2 M_2^2 -\t{\alpha}_1 M_1^2 \nonumber\\
\frac{dm_D^2}{dt}&=&\frac{16}{3}\t{\alpha}_3 M_3^2 +\frac{4}{9}\t{\alpha}_1 
M_1^2
\eeqa
and
\beq
\frac{d\mu_2^2}{dt}=3\t{\alpha}_2M_2^2 +\t{\alpha}_1^2 M_1^2 +
(3\t{\alpha}_2+\t{\alpha}_1)\mu^2-6Y_t\mu_2^2-3Y_t\left( A_t^2m^2_{3/2}-\mu^2+
\frac{1}{2}m_D^2-\frac{1}{2}m_4^2-\frac{3}{2}m_5^2 \right).
\eeq
 
\section{Solutions of the renormalisation group equations}
Analytic solutions of the equations of appendix A may be obtained along the
lines of refs.\cite{ibanez3} and \cite{ibanez4}. Including the contribution of
the $\mu$ parameter they are as folllows.
\beqa
\t{\alpha}_i(t)&=& \t{\alpha}_i(0)(1+\beta_it)^{-1} \\
\frac{M_i(t)}{M_i(0)}&=&\frac{\t{\alpha}_i (t)}{\t{\alpha}_i (0)} \\
Y_t(t)&=&\frac{Y_t(0)E(t)}{1+6Y_t(0)F(t))} \label{Yt}
\eeqa
where
\beq
F(t)=\int^t_0 E(t') dt' 
\eeq
Also,
\beq
\mu^2(t)=\mu^2(0) \frac{ (1+\beta_2t)^{\frac{3}{b_2}} 
(1+\beta_1t)^{\frac{1}{b_1}} }{(1+6Y_t(0)F)^{\frac{1}{2}}} 
\equiv \mu^2(0)q^2(t)  
\eeq
\beqa
m_{3/2}(1+6Y_t(0)F(t))A_t(t)-m_{3/2}A_t(0)&=&(1+6Y_t(0)F(t)) 
\left( \frac{16}{3}\t{\alpha}_3(0)M_3(0)h_3(t)\right. \nonumber\\ 
& &+\left.3\t{\alpha}_2(0)M_2(0)h_2(t)+
\frac{13}{9}\t{\alpha}_1(0)M_1(0)h_1(t) \right) \nonumber\\
& &-6Y_t(0)I(t)
\eeqa
where
\beq 
h_i(t)=\frac{t}{(1+\beta_it)}
\eeq
and
\beqa
I(t)&=&\int^t_0dt't'(1+\beta_3t')^{\frac{16}{3b_3}}(1+\beta_2t')^
{\frac{3}{b_2}}
(1+\beta_1t')^{\frac{13}{9 b_1}}    \nonumber\\
& &\times\left[\frac{\frac{16}{3}\t{\alpha}_3(0)M_3(0)}{(1+\beta_3t')} +
\frac{3\t{\alpha}_2(0)M_2(0)}{(1+\beta_2t')} +\frac{\frac{13}{9}\t{\alpha}_1
(0)M_1(0)}{(1+\beta_1 t')} \right].
\eeqa
The corresponding $A$ terms for the first two generations are obtained by 
putting $Y_t(0)=0$.
\beqa
B(t)-B(0)&=&\frac{1}{2}(A_t(t)-A_t(0)) \nonumber\\
& &+m^{-1}_{3/2}\left(\frac{-8}{3}\t{\alpha}_3(0)M_3(0)h_3(t)+\frac{3}{2}
\t{\alpha}_2(0)M_2(0)h_2(t)+\frac{5}{18}\t{\alpha}_1(0)M_1(0)h_1(t)\right).
\nonumber\\
\eeqa
For the first and second generation supersymmetric partners of quarks and 
leptons
\beq
m_\alpha^2(t)-m_\alpha^2(0)=2\sum^3_{i=1}C^\alpha_i\t{\alpha}_i(0)M_i^2(0)
f_i(t)
\eeq
where
\beq
f_i(t)=\beta_i^{-1}(1-(1+\beta_it)^{-2}) \ .
\eeq
Also for the masses $\mu_1$ and $\mu_3$ in the Higgs scalar potential (the 
equation for $\mu_2$ is integrated numerically),
\beq
\mu_1^2(t)=\mu_1^2(0)-\mu^2(0)+\mu^2(0)q^2(t)+\frac{3}{2}\t{\alpha}_2(0)
M_2^2(0)f_2(t) +\frac{1}{2}\t{\alpha}_1(0)M_1^2(0)f_1(t) 
\eeq
and
\beqa
\mu_3^2(t)&=&q(t)\mu_3^2(0)+\frac{3q(t)Y_t(0)}{(1+6Y_t(o)F(t))}A_t(0)m_{3/2}
\mu(0)-\mu(0)q(t)(3\t{\alpha}_2(0)M_2(0)h_2(t) \nonumber\\
& &+\t{\alpha}_1(0)M_1(0)h_1(t)) +\frac{3Y_t(0)\mu(0)q(t)I(t)}
{(1+6Y_t(0)F(t))}\ \  .
\eeqa
Finally for the masses of the supersymmetric partners of the third generation
quarks and leptons
\beqa
m_4^2(t)-m_4^2(0)&=&-3\t{\alpha}_2(0)M_2^2(0)f_2(t)+\t{\alpha}_1(0)M_1^2(0)
f_1(t) \nonumber\\
m_5^2(t)-m_5^2(0)&=&-\frac{8}{3}\t{\alpha}_3(0)M_3^2(0)f_3(t)+\t{\alpha}_2(0)
M_2^2(0)f_2(t)-\frac{5}{9}\t{\alpha}_1(0)M_1^2(0)f_1(t) \nonumber\\
m_6^2(t)-m_6^2(0)&=&4\t{\alpha}_3(0)M_3^2(0)f_3(t)+\frac{1}{3}\t{\alpha}_1(0)
M_1^2(0)f_1(t) \nonumber\\
m_7^2(t)-m_7^2(0)&=&\frac{3}{2}\t{\alpha}_2(0)M_2^2(0)f_2(t)-\frac{1}{2}
\t{\alpha}_1(0)M_1^2(0)f_1(t) \nonumber\\
m_D^2(t)-m_D^2(0)&=&\frac{8}{3}\t{\alpha}_3(0)M_3^2(0)f_3(t)+\frac{2}{9}
\t{\alpha}_1(0)M_1^2(0)f_1(t) \ .    
\eeqa

\end{document}